\documentclass[conference]{IEEEtran}
\IEEEoverridecommandlockouts
\usepackage{cite}
\usepackage{amsmath,amssymb,amsfonts}
\usepackage{algorithmic}
\usepackage{graphicx}
\usepackage{textcomp}
\usepackage{xcolor}
\usepackage{hyperref}
\usepackage{tabularx}
\def\BibTeX{{\rm B\kern-.05em{\sc i\kern-.025em b}\kern-.08em
    T\kern-.1667em\lower.7ex\hbox{E}\kern-.125emX}}
\begin{document}


\title{SkillScope: A Tool to Predict Fine-Grained Skills Needed to Solve Issues on GitHub\\

}

\author{\IEEEauthorblockN{
Benjamin C. Carter\textsuperscript{$1$},
Jonathan Rivas Contreras\textsuperscript{$1$}, 
Carlos A. Llanes Villegas\textsuperscript{$1$},
Pawan Acharya\textsuperscript{$2$}, \\
Jack Utzerath\textsuperscript{$1$},
Adonijah O. Farner\textsuperscript{$1$},
Hunter Jenkins\textsuperscript{$1$}, 
Dylan Johnson\textsuperscript{$1$},
Jacob Penney\textsuperscript{$2$}, \\
Igor Steinmacher\textsuperscript{$2$},
Marco A. Gerosa\textsuperscript{$2$},
Fabio Santos\textsuperscript{$3$}}
\vspace{1em}
\IEEEauthorblockA{\textsuperscript{$1$}\textit{Grand Canyon University, USA, \textsuperscript{$2$}Northern Arizona University, USA, \textsuperscript{$3$}Colorado State University, USA}}

BCarter44@my.gcu.edu, JRivas20@my.gcu.edu, CLlanesVil@my.gcu.edu, pawan\_acharya@nau.edu, jutzerath1@my.gcu.edu \\ afarner@my.gcu.edu, hjenkins14@my.gcu.edu ,
DJohnson457@my.gcu.edu,jacob\_penney@nau.edu, \\igor.steinmacher@nau.edu, marco.gerosa@nau.edu, fabio.deabreusantos@colostate.edu
\vspace{-1em}
}

\maketitle

\begin{abstract}
New contributors often struggle to find tasks that they can tackle when onboarding onto a new Open Source Software (OSS) project. One reason for this difficulty is that issue trackers lack explanations about the knowledge or skills needed to complete a given task successfully. These explanations can be complex and time-consuming to produce. Past research has partially addressed this problem by labeling issues with issue types, issue difficulty level, and issue skills. 
However, current approaches are limited to a small set of labels and lack in-depth details about their semantics, which may not sufficiently help contributors identify suitable issues. To surmount this limitation, this paper explores large language models (LLMs) and Random Forest (RF) 
to predict the multilevel skills required to solve the open issues. We introduce a novel tool, \textit{SkillScope}, which retrieves current issues from Java projects hosted on GitHub and predicts the multilevel programming skills required to resolve these issues. In a case study, we demonstrate that \textit{SkillScope} could predict 217 multilevel skills for tasks with 91\% precision, 88\% recall, and 89\% F-measure on average. Practitioners can use this tool to better delegate or choose tasks to solve in OSS projects.
\end{abstract}

\vspace{-1.2em}
\section{Introduction}
\label{sec:introduction}

Newcomers to Open Source Software (OSS) projects struggle to find suitable tasks, hindering the integration of contributors into communities that require their contributions to survive~\cite{steinmacher2015systematic}. Revealing issue types and skills needed to solve issues would help newcomers decide on where to contribute or maintainers where to allocate resources. 

Researchers and OSS project maintainers have proposed strategies to classify tasks and support newcomers~\cite{santos2022how}. 
Santos et al.~\cite{santos2023tell,santos2023tag}, for example, analyzed APIs as proxies for skills and trained machine learning models to predict the skills required to solve new tasks. Vargovich et al. designed a tool implementing a skill labeling approach~\cite{joe2023give} using BERT and RF based on 31 API domains. 
However, the API domains cover only high-level ``flat'' skills (e.g., ``UI'', ``IO'', ``Cloud'') and lack clarifying details about the domains they define. 
Moreover, each domain may encompass many subsumed tasks. For example, the ``Database'' domain can be divided into subdomains like ``Query Execution'', ``DB Security'', and ``DB Backup''. A practitioner does not necessarily excel in all subdomains to solve an issue where specific knowledge is needed. 
In addition, previous works \cite{santos2021can, santos2023tag} are constrained by semi-automatic processes, limited classification labels (31), and complex data pipelines to register the projects before allowing predictions \cite{joe2023give}. 

To address these problems, we introduce \textit{SkillScope}, 
a fully automatic skill label tool intended to increase skill specificity through multilevel classification. \textit{SkillScope} mines a repository's open issues and uses models trained on Abstract Syntax Trees (ASTs) of the source code to predict domains 
and subdomains relevant to each issue. The tool also offers a user interface (UI) to present the issues labeled with their corresponding predictions.

Like Santos et al.~\cite{santos2023tag}, we use API domains as a proxy for skills and employ an RF model to classify open issues based on a training set of closed issues and source code from various repositories. Complementing previous works, we investigate whether Large Language Models (LLMs) can handle 200+ multilevel API-domain labels.

Large Language Models (LLMs) have demonstrated strong capabilities in NLP and data classification tasks~\cite{colavito2024leveraging,aracena2024applying}, even though their performance can vary by task and domain~\cite{zhu2023labels, ziems2024can}, prompt~\cite{brown2020language}, and the context size provided~\cite{liu2024lost}. To assess how LLMs perform in classifying issues across different domains and subdomains, \textit{SkillScope} uses fine-tuned GPT models to classify issues across different domains and subdomains while comparing the performance to RF model.

Overall, \textit{SkillScope} successfully mines required data from repositories, categorizes the data into relevant skills, and allows users to make informed decisions about which issues to tackle. Through this, the tool seeks to strengthen the OSS ecosystem by providing easier entry to newcomers and allowing for faster issue-to-production for seasoned contributors. \textit{SkillScope} overcame the similar tools with a precision of 91\%.





\section{Related Work}\label{sec:related}

Other onboarding assistance tools were the subject of researchers' attention. Guizani et al. \cite{guizani2022attracting} organized repository data into a dashboard, offering maintainers a data fusion interface to attract and retain newcomers. Tools such as Ticket Tagger focused on labeling issues, helping to classify and organize them more efficiently\cite{kallis2019ticket,zhu2019bug}. 

Previous studies also showed how knowing the required skills can benefit maintainers and new contributors. Serrano et al. \cite{serrano2021find} proposed a chatbot to filter issues through pre-defined labels to help contributors find tasks based on their skills. Liang et al. \cite{liang2022understanding} created a model to understand which skills are used and the contributor's behavior toward them by mining signals that led to 45 unique skills, but it is not available as a tool for users. Santos et al. \cite{santos2021can, santos2023tag, santos2023tell} designed a method to predict and evaluate 31 API-domain skills, enabling the labeling of issues to assist with newcomer onboarding. Vargovich et al. \cite{joe2023give} tool selects issues based on users' skill selection in a UI using a semi-automatic categorization into the 31 domains available during the registration process.  Experts review and manually categorize around 8-45\% of the APIs that receive low similarity metrics. Collavito et al. \cite{colavito2024leveraging} investigated the extension to its LLMs (SEFIT and GPT-3.5) predicts issue types and found 83\% precision on average.

In contrast with previous works, \textit{SkillScope} offers RF, GPT-3.5, or GPT-4o-mini to categorize API-domain labels in two levels. We reused the 31 API domains proposed by Santos et al. \cite{santos2023tag} and identified an additional 186 subdomains by determining the relationship between libraries and methods. A maximum of 217 domains and subdomains can be achieved by automatically analyzing programs' AST trees without manual expert reviews. 
We fill the existing gap in tagging issues based on flat domains, providing more precise information to contributors about the skills needed to solve an issue. We also explore using large language models (LLMs) in this context.

\section{Tool Architecture}

\begin{figure}[htb]
\centering
\vspace{-10pt}
\includegraphics[width=0.48\textwidth]{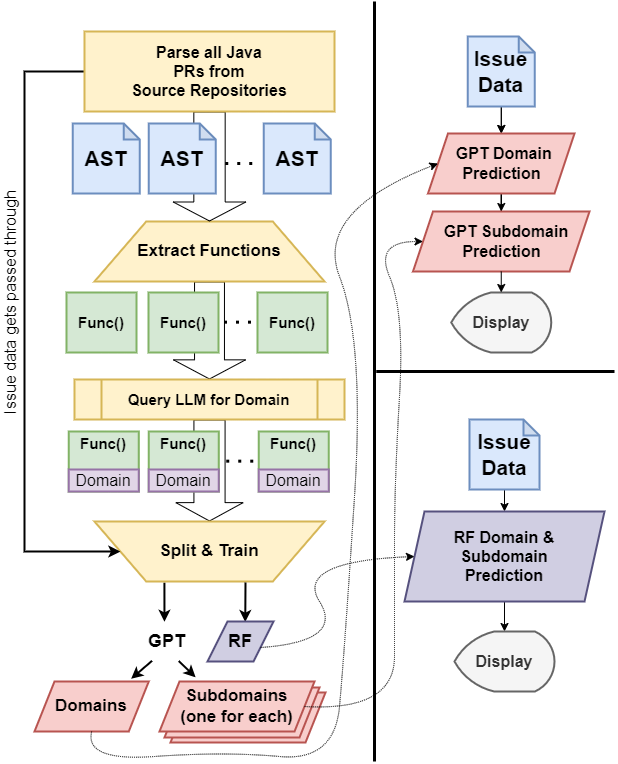}

\label{fig:architecture}

\caption{SkillScope Architecture}
\vspace{-10pt}
\end{figure}

\textit{SkillScope} has two workflows, training and issue classification. This is enacted by the four main components of the tool: data mining, the parsing engine, predictions, and user interface. The following sections will describe in detail what each component does in relation to the two workflows (Fig 1). 


\subsection{Data Mining} 
Before the tool is ready for predictions, it must be trained to recognize the skills required for an issue. To do this, the tool mines data from a sample of 11 source repositories stretching multiple fields, which 
are active and have over 3,000 stars (see replication package). The data extracted contains the issue text of closed issues and the respective source code related to each closed issue. Specifically, closed and merged PRs are considered to prevent PRs that were declined. Accordingly, \textit{SkillScope} 
collects all required data through endpoints exposed by the GitHub REST API v3 and feeds it to the parsing engine.



\subsection{Parsing Engine} 
 

The engine extracts each source file from the issue in the dataset. 
%
Each source file is parsed into an AST using tree-sitter-java\footnote{\url{https://github.com/tree-sitter/tree-sitter-java}}. From this, the engine determines the functions and classes used in the Java source files.
The functions and class names are first sent through zero-shot prompting into a tiered GPT-4o-mini prompt, first prompting for the domain and then prompting for the subdomain given the domain. The class name corresponds to the API domain, while the function name corresponds to the subdomain.
To prevent hallucinations, the output from the GPT-4o-mini model is fed into a similarity function using Spacy\footnote{\url{https://spacy.io/}} which evaluates the most-likely response given the list of domains and subdomains.

After, each source file in an issue is classified into domains/subdomains, which are combined to give a baseline for the classification of the entire issue. 
The final dataset holds columns for each possible domain and subdomain in a one-hot encoded format per issue, appended to the original issue data from the data mining process and stored in a database. 

Overall, eleven repositories
 across multiple fields were mined, with a combined 7245 pull requests extracted. Of these pull requests, about 57,206 Java source files were downloaded and processed. This resulted in 13,097 classes and 59,644 methods being categorized into 217 domains and subdomains.


\subsection{Predictions}
A one-vs-all classification strategy was utilized where predictions for domains and subdomains were separated~\cite{mirza2013one}. This resulted in two models for each domain: one model to identify the domains relevant to the issue, and another to determine which subdomains apply to the issue, in addition to the baseline RF model as earlier utilized by Santos et al. \cite{santos2023tag} for comparison. This approach was taken after tuning a single-layered GPT-3.5 model, which demonstrated poor performance due to prompt complexity and label imbalances.   

This process involved three steps: synthetic data creation, model fine-tuning, and evaluation.

\paragraph{Synthetic Data Creation}

For the GPT-4o-mini model, we created synthetic data to balance the minority domains/subdomains using LLM calls~\cite{guoa2023generative}, which proved to improve LLMs performance~\cite{gholami2023does} due to data scarcity and long tail for classification tasks~\cite{maheshwari2024efficacy}. 
We built individual positive instances data frames for each domain. Then, in the data frames with lower-than-average positive instances compared to the rest of the domains, synthetic data was created via OpenAI's GPT-4o-mini, which rephrased the titles and text of each issue. 
Since OpenAI API calls do not retain memory 
, the same client was used without the training data inadvertently leaking into the fine-tuning process \cite{maheshwari2024efficacy}. The GPT-3.5 model did not use any balance technique.

Once the data frames were created, each issue was cleaned following the process described by Aracena et al. \cite{aracena2024applying}, i.e., removing emojis, URLs, HTML tags, and other noise commonly found in GitHub text.

To handle the underrepresented domains with RF, MLSMOTE (Multi-Label Synthetic Minority Oversampling Technique) was leveraged to generate synthetic samples for the minority classes mimicking~\cite{santos2023tag}. 
The augmented dataset was then split into training and testing sets using an 80/20 split \cite{santos2023tag}. The RF model was trained on this data. 

We observed a substantial improvement in model performance with MLSMOTE. 
Micro metrics significantly improved, with a 38\% gain in precision. 

\paragraph{Training and Model Fine Tuning}

To train with the RF model, we used TF-IDF-Vectorizer to transform the textual data into a numerical format. 
According to the study by Cahyani \& Patasik\cite{cahyani2021performance}, TF-IDF fairs better than newer methods like Word2Vec or Doc2Vec when dealing with small, unstructured corpora-like issues and PRs in GitHub.

For the GPT4o-mini model, each domain was split with a 70 percent training and 30 percent testing set. Then, each training set was used to fine-tune a GPT-4o model following the process described by Aracena et al. \cite{aracena2024applying} with slight modifications. In fine-tuning, the input messages consisted of the domain description and the issue title and description, with a prompt asking the model to classify the issue. For output messages, they are 1 (domain is relevant to the issue) or 0 (domain is not relevant to the issue). The model was then fine-tuned from these messages, determining the suitability of a domain for an issue. Then, a separate model was trained for the subdomains from the same data, applying the same methodology but replacing the output message with the applicable subdomain. This left two models for each domain. First is one that is a binary classifier for the domain, and the second is a subdomain classifier. Both models are fine-tuned GPT-4o-mini models using a batch size of 1, a temperature of 1.0, and 3 epochs. These hyperparameters were found after experimentation for the best results. To prevent hallucinations in the subdomain model, the output from the GPT-4o-mini models is fed again into similarity functions using Spacy. GPT-3.5 model was trained mimicking \cite{aracena2024applying} in one single layer.

We also explored ExtremeML\cite{wydmuch2018no}, a more advanced model based on fastText designed for large-scale, imbalanced multilabel classification tasks. ExtremeML employs hierarchical softmax and negative sampling, making it highly efficient for handling numerous labels and imbalanced datasets. However, the performance was behind expectations, with only precision = 0.480, recall = 0.543, and F-measure = 0.345. Thus, we decided not to add it to the tool.

\subsection {Evaluation}
\label{sec:Evaluation}After training, we make predictions and calculate various performance metrics, including 
precision, recall, and an F1 score. 
The metrics for each label can be calculated using different averaging strategies. Since micro-averaging was used in previous studies \cite{santos2021can, santos2023tag} to calculate the predictions' metrics, we used it for the ability to provide comparisons. 


\subsection{User Interface}\label{sec:userinterface}

The UI for this tool uses Django, a web development framework. After filling out the form with the link to a Java project repository, the user may select how many issues and skills the tool must exhibit in return, as well as the algorithm. 
\textit{SkillScope} is available online \footnote{\url{https://skillscope.codingcando.com/}}. 
To promote reproducibility, \textit{SkillScope} replication packages are available. \footnote{\url{https://zenodo.org/records/14715839}} (UI and Engine) and \footnote{\url{https://doi.org/10.5281/zenodo.14715790}} (Training and Predictions).  












\section{Tool Evaluation}\label{sec:results}

To evaluate tool performance, we mined JabRef project contributions to compare with previous studies~\cite{joe2023give,santos2021can, santos2023tag}. 
\begin{figure}[htb]
\centering
\vspace{-10pt}
\includegraphics[width=0.5\textwidth]{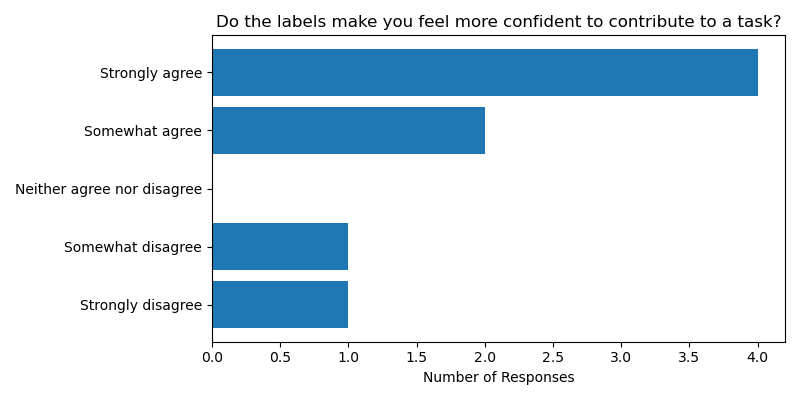}
\caption{Evaluation}
\label{fig:evaluation-likert}
\vspace{-4pt}
\end{figure}


The RF model surpassed both LLM models in predicting the domains and subdomains, as detailed in Table \ref{table:performance}. This performance also overcame Santos et al. \cite{santos2021can, santos2023tag}. The prediction speed of the RF model was also much faster than GPT-4o-mini, as it makes a single call per domain and then a secondary call for its subdomain. With RF, only one prediction is necessary. We also evaluated the user experience with contributor candidates. The survey was distributed via mail lists and social media. The survey saw eight respondents, with the majority finding that the tool either strongly or somewhat confident to contribute to open source projects in Fig \ref{fig:evaluation-likert}. Majority participants had little OSS contribution experience but 5+ years of programming experience. The replication package contains plots for other questions.

\label{table:performance}


\begin{table}[]
\centering
\begin{tabular}{llll}
Results                                         & Precision      & Recall         & F-1           \\ \hline
SkillScope RF + TF-IDF                          & \textbf{0.908} & \textbf{0.876} & \textbf{0.889} \\
Santos et al. RF+TF-IDF *3 \cite{santos2023tag}    & 0.864         & 0.786 &  0.810 \\
Santos et al. RF+TF-IDF \cite{santos2023tell}   & 0.842          & 0.835          & 0.838          \\
Collavito et al. *4 \cite{colavito2024leveraging}  & 0.832      & 0.832 & 0.832 \\
Vargovich et al. RF+TF-IDF \cite{joe2023give}   & 0.806          & 0.782          & 0.793          \\
Vargovich et al. BERT \cite{joe2023give}        & 0.791          & 0.606          & 0.686          \\
SkillScope GPT-3.5 LLM**                        & 0.756          & 0.214          & 0.335          \\
Santos et al. RF+TF-IDF \cite{santos2021can}    & 0.755          & 0.747          & 0.751          \\
SkillScope GPT-4o-min LLM*                      & 0.735          & 0.735          & 0.735          \\
Santos et al. BERT \cite{santos2023tag}      & 0.616          & 0.592          & 0.596          \\ \hline
\multicolumn{4}{l}{*all classes **class = 1 *3 projects avg *4 best micro avg}                                 
\end{tabular}
\label{table:performance}
\vspace{-13pt}
\end{table}



\section{Conclusion}
New contributors and maintainers face difficulty finding and assigning suitable tasks. We presented \textit{SkillScope}, a tool to reveal multilevel skills needed to solve open issues to mitigate these difficulties. SkilScope extended Santos et al. \cite{santos2021can, santos2023tag} studies and employs LLM and an RF model to predict 217 multilevel skills with a maximum precision of 0.908. 

Predicting fine-grained multilevel domains and subdomains gives the contributor more context about the skills involved in the task solution, which may decrease the confirmation bias. 

Future work should increase the number of domains and subdomains to represent the skills available in job market through applying extreme multilabel techniques. 
Also, parsing capability for other languages besides Java would increase the breadth of the model, as the current engine relies on tree-sitter-java for Java AST generation. In addition, we will use more repositories for training the models used by \textit{SkillScope}. The current model uses 11 repositories, but with more repositories, the model would be able to be more versatile to a variety of projects present in Open Source development. Santos et al. \cite{santos2021can, santos2023tag} evaluated the API domains with students and practitioners from the industry. \textit{SkillScope} will be evaluated in an empirical study with contributors to understand the role of the subdomains in choosing and solving tasks while adjusting the 186 subdomains scope. Extending the tool to mine the history of contributors will enable matching tasks with contributors' multilevel skills, enhancing the tool's applicability and allowing automatic recommendations. 
Finally, employing Retrieval-Augmented Generation to increase LLM metrics and, machine learning explanation techniques to ground users about how different skills were identified will be explored . 



\bibliographystyle{IEEEtran}
\bibliography{bibfile}

\end{document}